# SURVEY ON QOE\QOS CORRELATION MODELS FOR MULTIMEDIA SERVICES

Mohammed Alreshoodi[1] and John Woods[2]

[1] School of Computer Science and Electronic Engineering, University of Essex, UK
`mamalr@essex.ac.uk`

[2] School of Computer Science and Electronic Engineering, University of Essex, UK
`woodjt@essex.ac.uk`

## ABSTRACT

*This paper presents a brief review of some existing correlation models which attempt to map Quality of Service (QoS) to Quality of Experience (QoE) for multimedia services. The term QoS refers to deterministic network behaviour, so that data can be transported with a minimum of packet loss, delay and maximum bandwidth. QoE is a subjective measure that involves human dimensions; it ties together user perception, expectations, and experience of the application and network performance. The Holy Grail of subjective measurement is to predict it from the objective measurements; in other words predict QoE from a given set of QoS parameters or vice versa. Whilst there are many quality models for multimedia, most of them are only partial solutions to predicting QoE from a given QoS. This contribution analyses a number of previous attempts and optimisation techniquesthat can reliably compute the weighting coefficients for the QoS/QoE mapping.*

## 1. INTRODUCTION

In the past, the network has been examined objectively by measuringa number of criteria to determine the network quality. This quantification is called the Quality of Service (QoS) of the network. The term QoS refers to the ability of the network to achieve a more deterministic behaviour, so data can be transported with a minimum packet loss, delay and maximum bandwidth. One should note that QoS does not consider the user's perception. Another technique which takes into account the user's opinion is called Quality of Experience (QoE). The QoE is a subjective metric that involves human dimensions; it ties together user perception, expectations, and experience of application and network performance.

Adopting a more holistic understanding of quality as perceived by end-users (QoE) is becoming a vibrant area of research. When a customer has a low quality service, the service provider cannot afford to wait for customer complaints. According to an Accenture survey [1], about 90% of users do not want to complain about a low quality service, and they simply leave the provider and go to another. Therefore, it is essential that the service provider has a means of continually measuring the QoE and improve it as necessary.

A variety of factors can affect the perceived quality, including network reliability, the content preparation process and the terminal performance. The QoS of multimedia streaming services over IP networks is determined by several interdependent parameters. Some of the parameters can be adjusted, such as bandwidth and image resolution, while others are not like packet loss rate and delay. These missing parameters must be considered in order to increase the end user's satisfaction. However, the user's satisfaction is not only influenced by QoS parameters,there are also subjective factors (QoE) such as user experience, user interest and user expectation.

A number of researchers employ different methods according to the media type (e.g. voice, video and image). For each media type, there are a varietyof measurement methods having different computational and operational requirements. In this paper, a review ofthe QoE /QoS correlation models will be undertaken tohighlight the challenge of identifying the quantitative relationship between QoS and QoE.The reminder of this paper is organized as follows; an overview background is presented in section 2. In section 3, we present several QoE/QoS correlation approaches. In section 4, we presentan analytical review of the discussed QoE/QoS correlation approaches. In section 5, a discussion is presented. Finally, some conclusions are given in section 6.

## 2. BACKGROUND

### 2.1. QoS and QoE Layers

Different solutions for QoS have been proposed at a variety of layers in the OSI seven layersModel. The two layers generally used for QoS are the application and network layers [2]. The Application Layer includes services that are provided by the application in order to achieve the required QoS. Application layer QoS is concerned with parameters such as frame rate, resolution, colour, video and audio codec type, etc. On the other hand, network Layer services are provided by devices such as switches and routers. The network layer considers parameters such as delay, jitter, and packet loss, etc. Definitions from different authors suggest that a perceptual pseudo-layer can be imagined above both these two layers, which is concerned with the end-user's experience (QoE) [2]. Some authors consider this pseudo-layer as an extension to the Application layer [3], whereasothers view the QoE as an extension of traditional QoS because QoE provides information regarding the delivered services from the user's viewpoint [4].

QoS at the Application Layer is driven by human perception. The human perception of video services is based on two characteristics: spatial perception and temporal perception. In terms of video coding, three techniques are used to achieve the compression; which are Intraframe, Interframe and Entropy coding techniques. The QoS at the Network Layer can be classified into two main types: prioritisation and resource reservation. Different mechanisms and solutions can be used to form the QoS at the Network Layer, such as, Differentiated Services (DiffServ) [5], Integrated Services [6] and Multi-Protocol Label Switching (MPLS) [7].

Figure 1show the schematic relationship between QoS and QoE, which is divided into three zones. When the QoS disturbance is less than zone 1, QoE has a high value, i.e. the user's appreciation is not affected. The QoE decreases, when the QoS disturbance reaches zone 2. Finally, when the QoS disturbance increases to zone 3, the QoE may fall, i.e. the user's appreciation will be highly affected and they may stop using the service altogether. Typically, when the QoS disturbance parameter increases, the QoE metric and user's perception of quality decrease [8].

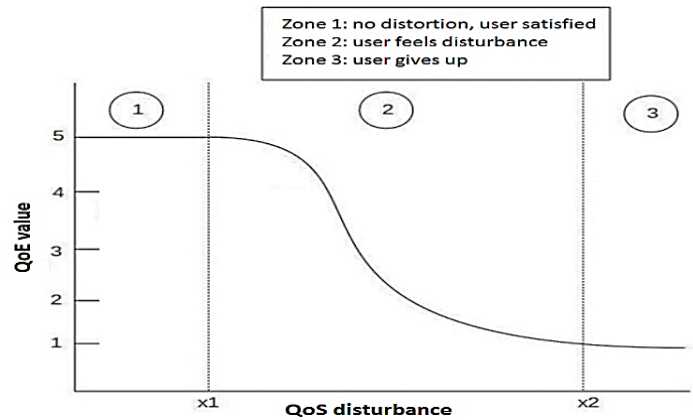

Figure 1: The mapping curve between QoE and QoS.

## 2.2.   QoE Measurement approaches

There are two main quality assessment methodologies, namely subjective and objective assessment. Measuring and ensuring good QoE of video applications is very subjective in nature. The most commonly used subjective method for quality measurement is the Mean Opinion Score (MOS). MOS is standardized in the ITU-T recommendations [9], and it is defined as a numeric value going from 1 to 5 (i.e. poor to excellent).The main drawbacks of this approach are: it is high in cost, time consuming, cannot be used in real time and lacks repeatability.These limitations have motivated the development of objective tools that predict subjective quality solely from physical characteristics.

By definition, the objective approach is based on mathematical and/or comparative techniques that generate a quantitative measure of the one-way video quality.This approachis useful for in-service quality monitoring or thedesign of networks/terminals, as well as in codec optimization and selection. An objective approach can be intrusive or non-intrusive. The intrusive methods are based on signals, while non-intrusive methods are based on network/application parameters. Generally, intrusive methods are accurate, but impracticable for monitoring live traffic because of the need for the original sequence, i.e. full reference quality measurement. Non-intrusive models do not require a copy of the original. Objective approachesusuallyignore the content type and the way the content is perceived by Human Visual System (HVS) [10]. For example, some objective methods try to compare original and received signals pixel by pixel to detect signal distortions, such as Peak Signal to Noise Ratio (PSNR) [11].

A combination of the objective and subjective approaches can be performed to overcome the shortcomings of each individual technique. The PSNR-mapped-to-MOS technique is a commonly adopted method that is used to estimate video QoE which has been affected by network conditions. This technique was demonstrated by several researchers to be inaccurate in terms of the correlation to the perceived visual quality [12] [13]. However, several modifications have been proposed to enhance the estimation accuracy. Several communities have ratified the improved PSNR-mapped-to-MOS technique, such as the ANSI T1.801.03 Standard [14] and the ITU-T in J.144 Recommendation [15].

Peter and Bjørn [16] classified the existing approaches of measuring network service quality from a user perspective into three categories, namely: (1) Testing User-perceived QoS (TUQ), (2) Surveying Subjective QoE (SSQ) and (3) Modelling Media Quality (MMQ), see Figure 2. The first two approaches collect subjective data from users, whereas the third approach is based on objective technical measurements. Based on this classification, we review the measurement approaches used for the selected models.

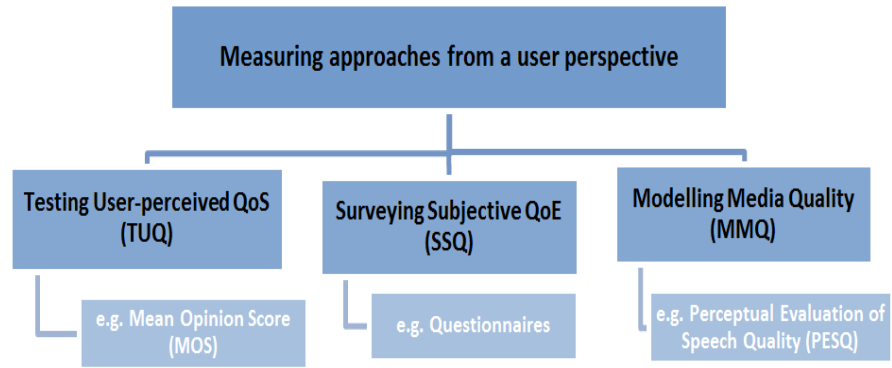

Figure 2: The approaches for measuring network service quality from a user perspective

## 2.3. Classification of Objective Quality Assessment Models

It is important to investigate the end user-oriented QoE versus network-oriented QoS parameters. This motivates the need to gain insight into the principal ways in which the quantitative parts of QoE are affected by the network QoS parameters. It is possible to measure and quantify the QoE and subsequently derive a mapping correlating the QoS parameters with the measured QoE metrics. Thus, it is possible to build an effective QoE-aware QoS model. A number of objective models have been devised for estimating QoE. The International Telecommunication Union (ITU) has developed a classification [17] to standardize these models based on a focus of each model type. Generally, the objective quality assessment methodologies can be categorized into five types:

- **Parametric packet-layer model** predicts QoE from packet-header information, without handling the media signal itself. It does not look at the payload information; therefore it has difficulty in evaluating the content dependence of QoE.
- **Parametric planning model** takes quality planning parameters for networks and terminals as its input. This type of model requires a priori information about the system under testing.
- **Media layer model** predicts the QoE by analysing the media signal via HVS. However, if media signals are not available, this type of model cannot be used.
- **Bit-stream model** is a new concept. Its position is in between the parametric packet-layer model and the media-layer model. It derives the quality by extracting and analysing content characteristics from the coded bit-stream.
- **Hybrid model** is a combination of some or all of these models. It is an effective model in terms of exploiting as much information as possible to predict the QoE.

## 2.4. Mapping Function

Since subjective scores and objective quality indices typically have different ranges, a meaningful mapping function is required to map the objective video quality (VQ) into the predicted subjective score (MOSp). Mapping functions can be categorised into linear and non-linear. The linear mapping function (1) can be used when both objective and subjective are scaled uniformly, i.e. an equal numerical difference corresponds to an equal perceived quality difference over the whole range [18].

$$MOS_p = a1 + a2 * VQ \qquad (1)$$

*a1* and *a2* parameters can be obtained by applying a linear fit between the VQ values and the respective MOS scores. After that, in order to evaluate the objective metric, MOSp values can be solved, as well asthe predicted scores which need to be correlatedanalytically to the actual scores.Nevertheless, the objective quality scalesare rarely uniform. The linear mapping function may also providea pessimistic view of the performance. The nonlinear mapping functions overcome this issue, and this is why they are widely used. Typically, the nonlinear mapping functions yield significantly higher correlations than the linear counterparts.The common mapping functions in the literature are Logistic (2), Cubic (3), Exponential(4), Logarithmic (5) and Power functions (6). These different forms of mapping function correspond to different QoS and QoE parameter measurements [18].

$$MOS_p = \frac{a1}{1 + exp\,[-a2 * (VQ - a3)]} \qquad (2)$$
$$MOS_p = a1 + a2*VQ + a3*VQ^2 + a3*VQ^3 \qquad (3)$$
$$MOS_p = a1 * exp\,(a2 * VQ) + a3 *exp\,(a4 * VQ) \qquad (4)$$
$$MOS_p = a1 - a2|\,log\,(VQ)\,| \qquad (5)$$
$$MOS_p = a1 *VQ^{a2} + a3 \qquad (6)$$

## 3.    QoS/QoE Correlation Approaches

From a literature review of QoE\QoS correlation models for multimedia services, a number of general modelling approaches can be identified and are analysed in this paper in order to obtain a "broader picture" on this topic. Some of the selected modelswill be discussed and reviewed in this section, while others aresummarised in tables.

### 3.1. IQX hypothesis

In [8], the QoE variation has been studied quantitatively with respect to QoS disturbances. The authors discuss the difficulty of matching the subjective quality perception to the objective measurable QoS parameters. They assumed a linear dependence on the QoE level, which implies the following differential equation:

$$\frac{\partial QoE}{\partial QoS} \sim -\,(QoE - \gamma)(7)$$

Equation (7) is solved by a generic exponential function called IQX hypothesis (8), which relates changes of QoE with respect to QoS to the current level of QoE.

$$QoE = \alpha * exp\,(-\,\beta * QoS) + \gamma \qquad (\alpha,\ \ \beta\ and\ \gamma\ are\ positive\ parameters) \qquad (8)$$

In this study, the QoE has been considered as theMOS, while the QoS has been evaluated through three different criteria: packet loss, jitter, response and download times. The IQX hypothesis is tested by only two different applications, VoIP and web browsing. The authors made a comparison between the IQX hypothesis and the logarithmic expression given in [19] in order to highlight the improvement. From the result, the proposed IQX approximations were of a better quality than the logarithmic.However, the main weakness of the proposed model is that it does not consider how the time-varying nature of IP (internet Protocol) impairments impacts the quality as perceived by the end user.Also, the used network emulator (NIST Net) [20] did not manage to create auto-correlated packetloss; this means that packets were dropped randomly.

Therefore,such studies require the incorporation of a different network emulator to emulate burst losses. Other works have also proposeda generic relationship between QoE and QoS with respect to a single QoS parameter in each experiment. Interested readers are invited to refer to[21], [22] and [23].

## 3.2.   VQM-based Mapping Model

The authors in [24] proposed an objective approach that provides a multidimensional QoS/QoE mapping hierarchy, in which the QoE is indicated by the Video Quality Metric (VQM) [25]. In this study, the VQM is a function of n-dimensional QoS (where n is the number of different QoS parameters), i.e.

$$VQM = f(X1, X2, X3, \ldots\ldots\ldots..Xn) \quad (X \text{ is QoS parameter}) \quad (9)$$

Three QoS parameters were consider in this study namely packet loss, delay and jitter. NTIA General Model [26] was used to produce the VQM scores. The emulation results have been analysed and a simple formula derived in order to predict the QoE from QoS for streaming video under the given conditions. The VQM function was expressed mathematically in the form given by equation (9) by using the curve fitting tool IstOpt [27], for n = 2. The final result was given by equation (10).

$$VQM = \frac{P1 + P2*X_1 + P3*X_1^2 + P4*X_2 + P5*X_2^2}{1 + P6*X_1 + P7*X_1^2 + P8*X_2 + P9*X_2^2} \, s.t. \, 0 \leq X1 \leq 5 \, , 0 \leq X2 \leq 5 \quad (10)$$

Where X1 denotes the jitter and X2 denotes the packet loss percentage. The variation of VQM score with the packet loss ratio was obtained from four video samples. The results illustrate that as the packets loss increases, the VQM score increases as well, which indicates that a worse QoE will result. Similarly, the dependency of the VQM score on delay and jitter, was recorded for thesame four video samples. Emulation results from other video samples were also used to verify equation (10). However, it is argued that equation (10) is not the best fit and insufficient to prove the causal QoE/QoS mapping. So, in order to have a better fit, more complex curve fitting algorithms are needed. This study focused only on the QoS parameters at the Network Layer. Another work [28] also proposed treating the relationship between the QoS and the VQM quality metric, which is considered an indicator of QoE.

## 3.3.   QoEModel usingStatistical Analysis method

In an effort to reduce the need for subjective studies, the authors in [29] present a method that only relies on limited subjective testing. Viewers were presented with the same video in a descending or ascending order of quality. Then, the viewers marked the point at which the change of quality became noticeable by using the method of limits [30]. Discriminate Analysis (DA) [31]was used to predict group membershipsfrom a set of quantitative variables. The group memberships were separated by mathematical equations and then derived.The derived equations are known as discriminant functions, which are used for prediction purposes. The general discriminant function formula is given in equation (11).

$$f_{km} = u_0 + u_1 X_{1km} + u_2 X_{2km} + \ldots\ldots + u_p X_{pkm} \quad (11)$$

$f_{km}$= predicted discriminant score for case $m$ in group $k$
$X_{ikm}$= value of the quantitative predictors for case $m$ in group $k$
$u_i$= Coefficients for variable $i$, for $i = 1 \ldots, p$

In this study, two video parametershave been used namely the bitrate and the frame rate for three different terminals and six types of video content.The authors explain that involving other factors related to the video content and coding parameters can maximize the user perceived quality and achieve efficient network utilization. The accuracy of the developed model validated for each terminal was Mobile phones: 76.9%, PDAs: 86.6% and Laptops: 83.9%.However, this approach suffers from a limited accuracy due to the statistical method used to build the prediction models. Moreover, no specific implementation was considered for the QoS parameters at the Network Layer.Another work [32] has also used statistical analysis and learning as a key for optimizing video QoE.

## 3.4.    QoEModels based onMachine Learning methods

The previous work in [29] was further extended in [33], where QoE prediction modelsarebuilt by using Machine Learning methods: Decision Trees (DT) [34] and Support Vector Machines (SVM) [35]. The developed models in this study perform with an accuracy above 90% estimated using the cross-validation technique [36]. From the results, the accuracy of the DT model (J84)is validated for three terminals and was:93.55% (Mobile phones),90.29% (PDA) and95.46% (Laptop). For the SVM model (SMO) the accuracy was 88.59% (Mobile phones), 89.38% (PDAs) and 91.45% (Laptops).They found from the results that both methods outperform the Discriminate Analysis method which was used in [29]. However, the error of these modelsis between 10% and 20%, as shown in [37].Work in [38] has also used DT and SVM for building an objective QoE model and then compares them with other Machine Learning methods including: Naive Bayes (NB), kNearest Neighbours (k-NN), Random Forest (RF) and Neural Networks (NNet). Results show that RF performs slightly better than the others examined.

The study reported in [37] proposed a method that connects the QoE metrics directly to QoS metrics according to the corresponding level of QoE over a WiMAX network. The QoE was estimated by employing a Multilayer Artificial Neural Network (ANN).The network QoS parameters were selected as the input layer, while the MOS, PSNR, SSIM and VQM as the output layer. The ANN model was trained to get the correct weights. A video database was utilized for the estimation and gathering of results. This database was divided into three smaller groups; 70% for ANN training, 15% for testing and 15% to validate the training process of the database. After training the ANN model, the relationship between the input layer and output layer was established. From the results, the proposed model gives acceptable prediction accuracy. The developed model can also adjust the input network parameters to get an ideal output to satisfy the users' need. Nevertheless, a larger database is needed to serve as an input to the neural network, and cope with all scenarios. Further the proposed model does not rely on any interaction from real humans, so it is not time consuming. Other works like [39], [40] and [41] also used the ANN method to adjust the input network parameters to get the ideal output to satisfy the users' need. Basically, the success of the ANN approach depends on the model's ability to fully learn the non-linear relationships between QoSand QoE.

The authors in [42] proposed two learning prediction models to predict the video quality in terms of MOS. The first model based on an Adaptive Neural Fuzzy Inference System (ANFIS) [43], while the second is based on a nonlinear regression analysis. They also investigate the impact of QoS on end-to-end video quality for H.264 encoded video and the impact of radio link loss models(2-state Markov models) over UMTS networks. A combination of physical and application layer parameters were used to train both models, and then validated with an unseen dataset. The results demonstrate that both models give good prediction accuracy. However, the authors concluded that the choice of parameters is crucial in achieving good prediction accuracy. The proposed models in this paper need to be validated by more subjective testing. Other works like [44] have also used the ANFIS approach to identify the causal relationship between the QoS parameters that affect the QoE and the overall perceived QoE.

## 3.5.    QoE model using Crowdsourcing for subjective tests

In [45], theauthors address the challenge of assessing and modelling QoE for online video services that are based on TCP-streaming, such as YouTube. A dedicated YouTube QoE model was presented that map between the user ratings and video stalling events that are caused by network bottlenecks. Also, a generic subjective QoE assessment methodology was proposed that is based on a Crowdsourcing approach, which is a highly cost-efficient and flexible way of conducting user experiments. Crowdsourcing means to outsource a task in the form of an open call to a crowd of anonymous users. The Microworkers1 crowdsourcing platform [46] was used to conduct online user surveys. Also, the SOS hypothesis [47] was used to analyse and filter the user diversity (e.g. individual expectations regarding quality levels, uncertainty how to rate, user type and sensitivity to impairments etc.) before the key influence factors on YouTube QoE are investigated.

The study's results indicate that QoE is primarily influenced by the video stalling events. In contrast, they did not detect any significant impact of other factors including internet usage level, age, or video content type. They found that users may tolerate one stalling event below three seconds per clip, but they tend to be highly dissatisfied with two or more stalling events.However, a lower level of reliability can be assumed due to the anonymity of users in the crowdsourcing platform, as well as loss of overall control. Some subjects may submit invalid or low quality work in order to maximize their received payment while reducing their own effort. Therefore, it is necessary to develop a sophisticated method to test the trustworthiness and quality of the test subjects. Work in [38] has also used the Crowdsourcing approach to collect QoE datasets.

## 3.6.    QoE model using a Resource Arbitration System

Reference [48] proposed a framework for estimating the QoE, based on the hypothesis that a better QoE can be achieved when all of the QoS parameters are arbitrated as a whole rather than looking at each of them individually. The Network and Application QoS metrics are weighted and represented by an overall single factor (QoE). The authors discussed that the measured QoE is a function of AQoS and NQoS, QoE = f (AQoS, NQoS). A resource arbitration system was designed in this study. Several experiments were also conducted using a QoS enabled network with feedback from software agents. Thus, users manually vary parameters at the application and network layers to achieve the QoE. The tool DSCQS (subjective testing) [49] was used to ask the end user to vote to show their pleasure or displeasure at the presentation. The PSNRs were compared both with and without using prioritization mechanisms. The QoE is increased by dynamically adjusting the three chosen network metrics: delay, jitter and packet loss. The research results illustrate that network arbitration improves the QoE. This implies that the interaction of both Network and Application Layer arbitration can provide a better QoE. Nevertheless, if improved video quality was observed, it was at the expense of other media services. The implementation of the proposed model focused only on the NQoS parameter.

## 3.7.    QoE model considering equipment and environment factors

In [50],a scheduling algorithm called QoE Based Scheduling (QBS) was proposed to make a more efficient use of the network resources. To achieve this goal, a QoE video model was developed that involvedboth environmental and equipment factors. Environment and equipment influence a users' QoE directly and profoundly with respect to the QoS in the network. The QoS parameters considered were throughput and bit rate.The basic idea in the proposed model is to test the hardware and environment parameters by the user's equipment itself. After that the demand for the network signals' quality will be adjusted according to the test's results, in order to meet the users' needs. From the results, they noticed that the users' satisfaction was improved byusing the QBS algorithm, especially for resource-constrained users.They found that the interference from the surrounding environment (e.g. light, noise, shaking, etc.) lead to different QoE levels for the user.So,a better QoE can be received by providing high-quality signals to high end devices with lowinterference environments and vice versa. Nevertheless, the authors argue that the proposed algorithm needs to be optimized further.Also, more in-depth tests of a users' QoE are required in various scenarios with the inclusion of further QoS parameters.

## 3.8. QoE model based on Quantitative and Qualitative Assessment

In [51] the authors adopt a combination of qualitative and quantitative approaches.Rough Set Theory (RST) [52] was used for quantitative assessment, whereas CCA framework (Catalog, Categorize, and Analyze) for qualitative assessment. The combined impact of AQoS and NQoS parameters and content characteristics over QoEwere emulated and analysed. With quantitative assessment, they learnt that the different types of video content require different levels of QoS support. Also the AQoS and NQoS parameters have different levels of impact over QoE.For qualitative assessment,they found that slow moving video clips got less negative comments than fast moving video clips. Moreover, variations in NQoS parameters causemore negative comments for both slow and fast moving video clips, while there was no severe trend in the video bit rate variation. It is obvious from the results that the quantitative data assessment matches with the overall trend in qualitative comments. Basically, the results may look quite intuitive because only four QoS parameters were considered in this study. However, in a real environment as the influencing factors increase, understanding the interdependence between them becomes very complex.

# 4. ANALYTICAL REVIEW

## 4.1. Evaluation Approaches on the Analysed Models

The following table helps to show which aspects are evaluated by givenmodel, as well as how many different aspects were evaluated in each model. The review of thesemodels helps us to compare and make conclusions from the findings.

Table 1: A summary of the selected evaluation approaches for each model

| Aspects / Model | AQoS Parameters | NQoS Parameters | Video coding Parameters | Subjective Metrics | Objective Metrics | Packets Prioritization | Artificial Adaptive Technique | Application | Simulation / Test bed | Technology | Reference Measurements |
|---|---|---|---|---|---|---|---|---|---|---|---|
| Fiedler et al [8] | ✕ | ✕ | | ✕ | | | | VOIP and WB | Simu/ test-b | Wired | RR |
| Siller et al [48] | | ✕ | ✕ | ✕ | ✕ | ✕ | | VS | test-b | Wired | FR |
| Wang et al [24] | | ✕ | | | ✕ | | | VS | Simu/ test-b | Wired | NR |
| Agboma et al [29] | ✕ | | | ✕ | | | | VS | test-b | Wireless | NR |
| Menkovski et al [33] | ✕ | | | ✕ | | | ✕ | VS | Simu/ test-b | Wireless | NR |

| | | | | | | | | | | |
|---|---|---|---|---|---|---|---|---|---|---|
| Machado et al [37] | | ✕ | | ✕ | ✕ | | ✕ | VS | Simu | WiMax | FR |
| Du et al [39] | | ✕ | | ✕ | ✕ | | ✕ | VS | Simu/test-b | Wired | FR |
| Kim et al [21] | | ✕ | | ✕ | | | | IPTV | Simu/test-b | Wired | NR |
| Khan et al [42] | ✕ | ✕ | | | ✕ | | ✕ | VS | Simu | Wireless | NR |
| Malinovski et al [44] | | ✕ | | ✕ | | | ✕ | VC | test-b | Wired | NR |
| Han et al [50] | ✕ | ✕ | | | ✕ | | | VS | Simu | Wireless | NR |
| Laghari et al [51] | ✕ | ✕ | | ✕ | | | | VS | test-b | Wireless | NR |
| Ramos et al [28] | | ✕ | ✕ | | ✕ | | | VS | Simu/test-b | Wired | NR |
| Koumaras et al [22] | | ✕ | ✕ | | ✕ | ✕ | | VS | Simu/test-b | Wired | NR |
| Frank et al [40] | ✕ | ✕ | ✕ | ✕ | | | ✕ | VS | Simu | Wired | NR |
| Calyam et al [41] | ✕ | ✕ | ✕ | ✕ | ✕ | | ✕ | IPTV | test-b | Wired | NR |
| Mok et al [23] | ✕ | ✕ | | ✕ | | | | VS | test-b | Wired | NR |
| Elkotob et al [32] | ✕ | ✕ | | ✕ | | | | VC | test-b | Wireless | NR |
| Mushtaq,et al [38] | | ✕ | ✕ | ✕ | | | ✕ | VS | Simu/test-b | Wired | NR |
| Hoßfeld et al [45] | ✕ | ✕ | | ✕ | | | | VS | test-b | Wired | NR |

**VS:** Video Streaming, **VC:** Video Conferencing, **VOIP:** Voice over IP, **WB:** Web Browsing
**FR** (Full reference): Both reference videos and outcome videos are required for the evaluation process
**NR**(No reference): only the outcome video is required for the evaluation process
**RR** (Reduced reference): extract some features of both (reference and outcome) and compare them.

### 4.1.1 Measurement Approaches

The classification of measurement approaches in [16] will be used in this analytical review. Most of the analysedmodelsused the TUQ approach to collect the subjective measurementsfrom users, [8, 21, 23, 32and 40]. In this context, the MOS scoreis the common method in the TUQ approach.Other models like [24, 28, 42 and 50] use the MMQ approach that is based on objective technical measurements (e.g. PSNR, VQM). The models in [22, 37, 39, 41 and48] measured both subjective and objective variables by using both TUQ and MMQ approaches. In [29, 33and 45], the SSQ approach was used which is based on surveying user opinion to collect the subjective measures as qualitative data. The models in [38, 44 and 51] collected the subjective measurements as qualitative and quantitative data, using both TUQ and SSQ approaches. According to the results from the literature, models that measure both subjective and objective variables in a quantitative way better reflect the complexity of QoE. Therefore, it is better to collect subjective measures as quantitative data rather than qualitative data. The main reason is that the quantitative data enables statistical descriptions and a combined analysis of subjective and objective variables for deriving global QoE measures [16].

Generally, there is a misunderstanding in the current literature that data from users is necessarily subjective, while objective measures can only be collected from technology. Nevertheless, it is possible to develop a test methodology with objective measures of user's behaviour, which is commonly considered as subjective. In this regard, task duration and number of mouse clicks are examples of objective measures of user performance [16].

### 4.1.2. Parameter Mappings

In the context of QoS/QoE correlation models, the QoS parameters are the coefficients of the non-linear function fits that are identified and tuned during the measurement cycle.The majority of QoE estimation models use a coefficient method to map the relationship between the input pattern (QoS parameters) and the output (QoE). This was found to be the case in the reviewedmodels.The model in [24] used a fixed coefficient, which means the curve fitting algorithm used does not provide the best fit and is only sufficient to prove the proposed concept.In contrast, the models in [8, 21, 22, 23, 28, 29, 32, 45, 48, 50 and 51]used an optimisation approach for the coefficientsbased on mapping the fitting functions. These mapping functionare able to find an optimal fit for the given measurement points. In these studies, they derived expressions via mathematical modelling of the dataset to calculate the QoE from the QoS parameters.

Other models in [37, 39, 40, 41, 42 and 44] useartificial optimisation techniques to learn the model and find an optimal fit QoE metric from the QoS parameters.There are several widely used artificial techniques (Machine Learning)used in this field, such as ANN, Fuzzy Logic (FL) and ANFIS. The ANN technique hasbeen used in [37, 39, 40 and 41], whereas the ANFIS technique in [42 and 44]. In all the reviewed models, the QoS parameters were selected as the inputs, while the MOS or objective metrics (e.g. PSNR, VQM, SSIM, etc.) were selected as the outputs. There is no reason why QoS to QoE cannot be reversible. These artificial techniques adjust the input network parameters to get the ideal output to satisfy the users' needs. However, the success of these adaptive techniques is based on the model's ability to fully learn the non-linear relationships between QoS parameters and QoE. Also from the results, a larger database is needed to serve as input to the Machine Learning model, with more solid values and a generalization of the network. Moreover, the choice of QoS parameters is crucial in achieving good QoE prediction accuracy. Figure 3 shows an example of video measurement framework with ANN system that was designed in [37].

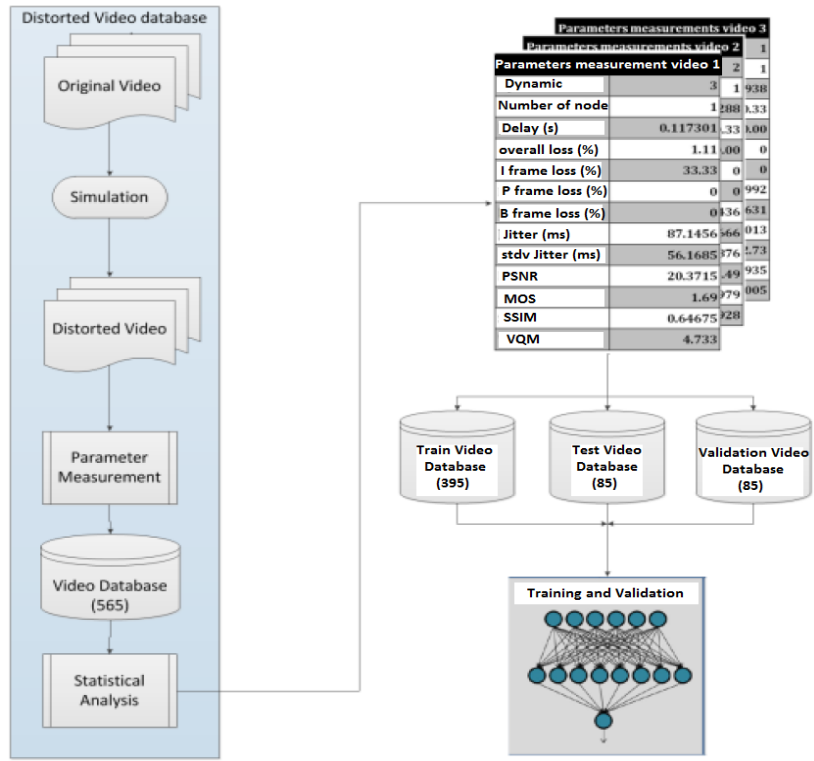

Figure 3: Video measurement framework with ANN system [37].

In addition, the authors in [33 and 38] used Machine Learning methodsas a techniquefor classifyingthecollected QoEdataset (which contains a set of labelled classes)to build the prediction model. After classifying the datasets, the ML method can make predictions about future instances. The ML can be categorised into; supervised and unsupervised learning methods. The supervised learning methods assume that the dataset's category structure is already known. In contrast, the unsupervised learning methods help to find the hidden structure in the unlabelled caseandthen classify it into meaningful categories.From the reviewed studies, the DT, SVM methods have been used in [33]as supervised learning classification because of the discrete nature of their datasets. In [38], the authors used theDT, SVM, NB, NNet, RF and k-NN methods as supervised learning classification and they made a performance comparison between these methods.

## 4.2. Selected QoE Factors onthe Analysed Models

Several communities and researchers proposed classifications of QoE influence factors into multiple dimensions.For example, the EU Qualinet community [53] groups QoE influence factors into three categories; (1) Human Factors (e.g., gender, age, education background, etc.), (2) System Factors (e.g., bandwidth, security, resolution, etc.), and (3) Context Factors (e.g., location, movements, costs, etc.). Also, Skorin-Kapov and Varela [54] categorized the QoE factors into four dimensions: Application, Resource, Context, and User.The following table summarises the selected factors affecting QoE on the analysedmodels.

Table 2: A summary of the selected factors that affect QoE on each model

| Model | Factors influencing Video QoE | | | | | | | | | |
|---|---|---|---|---|---|---|---|---|---|---|
| | Network impairments | Video content type | Video Quality | Coding parameters and components | Terminal device | QoS Mechanism | Home network characteristics | Environment | Human Factors | Time |
| Fiedler et al [8] | ✕ | | | | | | ✕ | | | ✕ |
| Siller et al [48] | ✕ | | ✕ | | | ✕ | | | | |
| Wang et al [24] | ✕ | ✕ | | | | | | | | |
| Agboma et al [29] | ✕ | ✕ | | ✕ | ✕ | | | | | |
| Menkovski et al [33] | ✕ | ✕ | | ✕ | ✕ | | | | | |
| Machado et al [37] | ✕ | ✕ | | | | ✕ | ✕ | | | |
| Du et al [39] | ✕ | | | | | | ✕ | | | |
| Kim et al [21] | ✕ | | ✕ | | | | ✕ | | | |
| Khan et al [42] | ✕ | ✕ | | | | | | | | |
| Malinovski et al [44] | ✕ | | | | | | ✕ | ✕ | ✕ | |
| Han et al [50] | ✕ | | ✕ | | ✕ | ✕ | ✕ | ✕ | | |
| Laghari et al [51] | ✕ | ✕ | | | | | | | ✕ | |
| Ramos et al [28] | ✕ | ✕ | | ✕ | | | | | | |
| Koumaras et al [22] | ✕ | ✕ | | ✕ | | | | | | |
| Frank et al [40] | ✕ | ✕ | | ✕ | | | | | | |
| Calyam et al [41] | ✕ | ✕ | ✕ | ✕ | | ✕ | | | | ✕ |
| Mok et al [23] | ✕ | | | | | | | | ✕ | |
| Elkotob et al [32] | ✕ | ✕ | | | ✕ | | ✕ | ✕ | | ✕ |
| Mushtaq et al [38] | ✕ | ✕ | ✕ | | | | | | ✕ | |
| Hoßfeld et al [45] | ✕ | ✕ | | | | | | | ✕ | ✕ |

The influence factor 'network impairments' was investigated by all the models. Along the transmission paths, various types of network impairments (e.g. loss, delay, jitter, etc.) occur due to the nature of IP-based networks. These network impairments can significantly impact the video quality. From the literature, the most evaluated NQoS parameters were packet loss, delay, and jitter, while bitrate and frame rate are used for AQoS parameters. Also, bandwidth and the congestion period were investigated in [21 and 39], whereas response and download time parameters were studied in [8]. In [42], Block Error Rate and Mean Burst Length were measured in the network layer. However, there are other QoS parameters that have rarely attracted attention including impact of burst of loss on perceived quality and prioritising the video packet depends on its type. Another interesting parameter for evaluation is the dejittering buffer size. Figure 4 show how the QoE is a non-linear function of QoS and influenced by many factors.

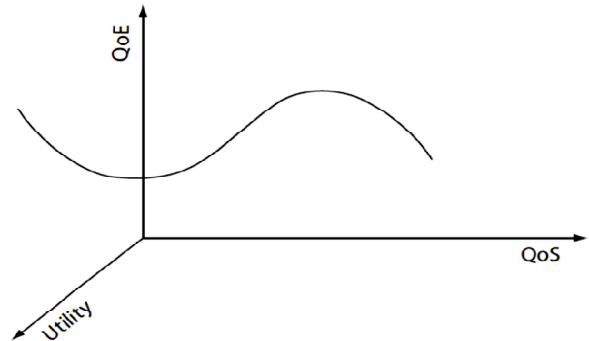

Figure 4: QoE is a non-linear function of QoS and many factors [55]

The second most studied QoE influence factor is 'video content type', covered by 14 studies.Video content has a significant effect on QoE. According to the study in [51], the main influencing factor for a slow moving video clip was packet loss alone, while for a fast moving football match clip, video bit rate as well as packet loss were important. With regard to video coding, the most used video coders in the reviewed studies were MPEG2, MPEG3, MPEG4, H.263 and H264. In contrast, layered video coding has rarely been considered. The H.264 Scalable Video Codec (SVC) [56] fits the requirements of video streaming in heterogeneous environments. It enables runtime-efficient scalability in three dimensions; spatial, temporal and fidelity. H.264/SVC performs a layered approach to achieve the scalability of the encoded video bit-stream.

There is however a number of parameters that affect the perceived quality and have rarely been investigated in the literature including:privacy concerns, interaction of audio and video, user interface, user's awareness of quality, cost andlast mile equipment/environment.In addition, quantization, (which means some amount of video information is thrown away during compression) introduces a certain amount of distortion into the video quality.Figure 5categories the reviewed studies based on the main QoE influence factors.

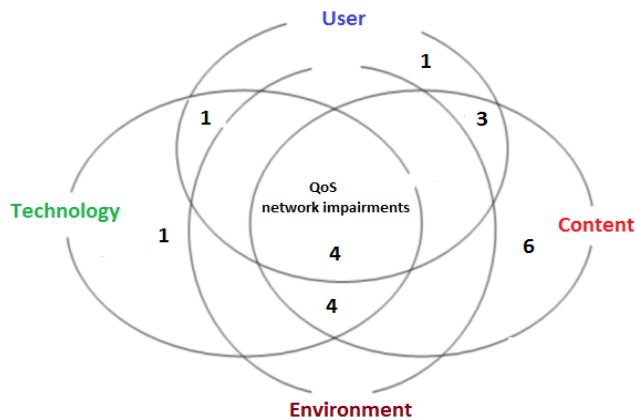

Figure 5: Categorising the reviewed studies basedon the main QoE influence factors.

### 4.3. The main findings of the studies analysed

The following table summaries the main findings of this literature survey to show what kind of information the authors were interested in.

Table 3: A summary of the main findings ofthe analysed models:

| Model | Results |
|-------|---------|
| Fiedler et al [8] | - It was proven that an exponential relationship exist between QoS and QoE.<br>- The proposed IQX hypothesis approximations were better quality than the logarithmic approximations |
| Siller et al [48] | - The network arbitration improves the QoE by using prioritization mechanisms.<br>- A better QoE can be achieved when all of the QoS parameters are arbitrated as a whole rather than looking at each of them individually |
| Wang et al [24] | - The VQM-based QoS/QoE mapping imports standardized VQM scores to provide mapping between QoS parameters and QoE in terms of perceived video quality. |
| Agboma et al [29] | - The developed model allows network operators to anticipate the user's experience and then allocate network resources accordingly.<br>- Model performs with accuracy of: Mobile phones 76.9%, PDA: 86.6%, Laptop: 83.9%. |
| Menkovski et al [33] | - The developed models are suitable for use in real world applications particularly as part of a control loop of a network management system.<br>- The models perform with an accuracy of:<br>J48model: Mobile phones: 93.55%, PDA: 90.29% and Laptop: 95.46%.<br>SMOmodel: Mobile phones: 88.59%, PDA: 89.38% and Laptop: 91.45%. |
| Machado et al [37] | - It was observed that the neural network for the scenario had a very good prediction.<br>- The influence of the video dynamics and the amount of nodes is perceptible, |
| Du et al [39] | - The method does not rely on interaction from real humans, and removes the need for human labour.<br>- BP Neural Network can adjust the input network parameters to get the ideal output to satisfy the users' needs. |
| Kim et al [21] | - The proposed model helps network providers to predict a subscriber's QoE in the provided network environment and analysesthe service environment |
| Khan et al [42] | - It is possible to predict the video quality if the appropriate parameters are chosen.<br>- The content type has a bigger impact on quality than the sender bitrate and frame rate.<br>- They found that the video quality is more sensitive to network level parameters compared to application level parameters. |
| Malinovski et al [44] | - The test experiment has confirmed the close relationship between QoS controls, the system's technical performance and the perceived subjective QoE of the viewers. |
| Han et al [50] | - Simulation results show that the average satisfaction is improved when the QBS algorithmis used, especially for the resource-constrained users.<br>- Considering user equipment and environment factors helps to achieve more efficient use of the network resource, and improves the average satisfaction of users. |
| Laghari et al [51] | - The different types of content require different level of QoS support. |
| | - Variation in NQoS parameters causes the generation of an abundant number of negative comments for video clips, while variation in bit rate does not have that such a severe trend.<br>- the overall trend in qualitative comments matches with quantitative data assessment |
| Ramos et al [28] | - The measurement workbench acquires both training data for model fitting and test data for model validation.<br>- Preliminary results show good correlation between measured and predicted values |
| Koumaras et al [22] | - It introduces the novel concept of predicting the video quality of an encoded service at the pre-encoding state, which provides new facilities at the content provider side.<br>- The proposed scheme can be applied on any MPEG-based encoded sequence, subject to a specific GOP structure. |
| Frank et al [40] | - From data analysis, they found that the usual parameters that can be controlled in an MPEG-4 codec do not have such a strong influence on the perceived video quality as a good network design that protects the video flows may do. |
| Calyam et al [41] | - The evaluation results demonstrate that the proposed models are pertinent for real-time QoE monitoring and resource adaptation in IPTV content delivery networks.<br>- Also, the model speed results suggest that over 10,000 flows can be handled in < 4ms. |
| Mok et al [23] | - The rebuffering frequency is the important factor responsible for the QoE variance<br>- Temporal structure, instead of spatial artifacts, is an important factor affecting the QoE. |
| Elkotob et al[32] | - The proposed scheme allows a mobile node to be proactively aware of the best access network for the next interval.<br>- This scheme would be interesting for operators and service providers who need to maintain graceful QoE profiles and optimize their resource usage. |
| Mushtaqet al [38] | - It is observed that DT has the best performance,in the case of mean absolute error rate, as compared to all other algorithms.<br>- After a statistical analysis of classification, the results show that RF performs slightly better than DT. |
| Hoßfeld et al [45] | - The proposed model indicates that users tolerate one stalling event per clip as long as stalling event duration remains below 3 s. In contrast, they did not detect any significant impact of other factors like level of internet usage, age or content type. |

## 5. DISCUSSION

It is obvious from the literature that the main objectives of the existing QoS/QoE correlation models are: (1) finding QoE with only knowledge of the QoS and (2) finding suitable QoS with a desired QoE. A number of researchers introduce different mapping methods for different media types (e.g. voice, video and image). For each media type, there are several methods of measurement that have quite different computational and operational requirements. In this context, discussing the efficiency of any QoE estimation model is difficult because there is lack of a standardisation.

There is also a lack of an accurate quantitative description of QoE. The interactions between various QoS parameters and their effects on QoE are still poorly understood, as well as there is no standardized methodology that directly and quantitatively maps QoS to QoE. Most of the current mappingmethods are only partial approaches to the QoE prediction issue. This motivates the need to introduce an efficient methodology for QoE\QoS correlation based on statistical descriptions and combinations of quantitative subjective measures with objective measures of the process and outcomes of usage. Generally, there are two main approaches for mapping the QoE/QoS relationship: A top-down approach, starting from the side of the user perspective (QoE), and a bottom-up approach starting from the network side (QoS). Also, a possible combination of the two approaches can be performed, see Figure 6.

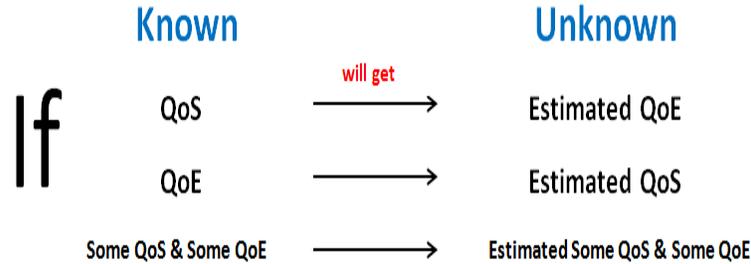

Figure 6: The mapping approaches

For parameter mappings, models that using intelligent optimisationtechniques outperform the other methods, as found in the literature. The artificial optimisation techniques are able to learn the model and find an optimal fit QoE metric from the QoS parameters.The success of these optimisation techniques is based on the model's ability to fully learn the non-linear relationships between QoSand QoE. For more accuracy, a larger video database is needed to serve as input to the model. Moreover, the choice of parameters is crucial in achieving good prediction accuracy. Also from the results, it is possible to develop a test methodology with objective measures of user's behaviour, which is commonly considered as subjective.

With regard to the challenges of QoE/QoS correlation modelling, there is a need to identify and understand many QoE influencing factors for a given type of service and how they influence each other. Some factors need to be carefully considered and may be difficult to measure. For example, a user's contextual information may help to provide contextualized QoE, but it raises some security and privacy issues.Furthermore, well planned quantitative subjective measurements need to be performed involving both cognitive and behavioural modelling. The test results need to be analysed to enable the derivation of key QoE influence factors and QoS parameters and their quality thresholds. This will helps to provide input for relevant QoE optimization strategies.

## 6.  CONCLUSION

Mapping the relationship between QoS and QoE is an extremely challenging task and the ultimate judge of multimedia service quality.End-to-end QoS is an important enabler for QoE and finding the correlation between them is a significant first step towards a more optimized feedback mechanism that is able to generate multimedia services in an efficient way. A number of QoE models have been selected and analysed in this paper in order to obtain a "broader picture" on the literature of QoE\QoS correlation models for multimedia services.We present how researchers undertook their studies by defining the evaluation approach on each study, such as the selected measurement approach, quality metrics, QoS parameters, etc. We also discussed the selected factors that affect QoE on each model.

From the literatures, most of the analysed models collect the subjective measures from users via the MOS score method. Some models measure both subjective and objective variables by using MOS and objective metrics. Othermodels have collected the subjective measures as qualitative data by surveying user opinion. In contrast, several models were only based on objective technical measurements. After comparing the studies' results, models that measure both subjective and objective variables in a quantitative way better reflect the complexity of the QoE. Therefore, it is better to collect subjective measures as quantitative data rather than qualitative data. The main reason is that the quantitative data enables statistical descriptions and a combined analysis of subjective and objective variables. Moreover, while most of the literature considers that objective measures can only be collected from technology, it is possible to develop a test methodology with objective measures of a user's behaviour, which is commonly considered as subjective.

In addition, the majority of the analysed models use a coefficient method to map the non-linear relationships between QoS parameters and QoE. According to the literature, models that useconstant coefficients will not provide the best fitfor the given measurement parameters.On the other hand, the optimisation approach to coefficients helps to find an optimal fit for the given measurement points. The studies that perform this method derived expressions via mathematical modelling of the dataset to calculate the QoE metric from the QoS parameters. In addition, other models usedartificial optimisation techniquesbased on ANN, Fuzzy Logic, DT, and SVM methods.These artificial techniques are able to adjust the input network parameters to get the optimal output to satisfy the users' need. However, the success of these artificial techniques is based on their ability to fully learn the causal relationship between input parameters (QoS) and the resulting QoE. In this regard, a larger database is needed to serve as input to the model. Moreover, the choice of parameters is crucial in achieving good prediction accuracy.

Overall, it becomes evident that while there are many QoE/QoS correlation models, most of them are only partial approaches to the QoE prediction issue. As such some models are too specific for a particular kind of application, as well as they have quite different computational and operational requirements. A model that gains insight into the principal ways in which the quantitative parts of QoE are affected directly by QoS parameters is still missing. This motivates the need to investigate the end user-oriented QoE versus network-oriented QoS parameters and find out meaningful mapping functions between them. In light of this there is a need to understand many QoE aspects and how they influence each other.